\documentclass[%
 reprint,
 amsmath,amssymb,
 aps,
 prd
floatfix,
showkeys
]{revtex4-1}

\usepackage[USenglish]{babel}
\usepackage{slashed}
\usepackage{physics}
\usepackage[normalem]{ulem}
\usepackage{graphicx}
\usepackage{dcolumn}
\usepackage{bm}
\usepackage[usenames]{color}
\usepackage{xcolor}
\usepackage[bf]{caption} 
\usepackage{float}
\usepackage{hyperref}
\usepackage{verbatim}

\newcommand{\be}{\begin{equation}}
\newcommand{\ee}{\end{equation}}
\newcommand{\bea}{\begin{eqnarray}}
\newcommand{\eea}{\end{eqnarray}}
\newcommand{\beas}{\begin{eqnarray*}}
\newcommand{\eeas}{\end{eqnarray*}}

\usepackage{tikz}
\usepackage{subcaption}
\newcommand\unity{1\!\!1}
\begin{document}

\title{QCD equation of state at finite isospin density from the linear sigma model with quarks: The cold case}


\author{Alejandro Ayala$^{1,2,3}$, Aritra Bandyopadhyay$^{3,4,5}$, Ricardo L. S. Farias$^3$, L. A. Hern\'andez$^{6,2}$, Jos\'e Luis Hern\'andez$^{1,7,8,9}$ }
\affiliation{%
$^1$Instituto de Ciencias Nucleares, Universidad Nacional Aut\'onoma de M\'exico, Apartado Postal 70-543, CdMx 04510, Mexico.\\
$^2$Centre for Theoretical and Mathematical Physics, and Department of Physics, University of Cape Town, Rondebosch 7700, South Africa.\\
$^3$Departamento de F\'isica, Universidade Federal de Santa Maria, Santa Maria, RS 97105-900, Brazil.\\
$^4$Guangdong Provincial Key Laboratory of Nuclear Science, Institute of Quantum Matter, 
South China Normal University, Guangzhou 510006, China.\\
$^5$Institut für Theoretische Physik, Universität Heidelberg, Philosophenweg 16, 69120 Heidelberg, Germany.\\
$^6$Departamento de F\'isica, Universidad Aut\'onoma Metropolitana-Iztapalapa, Av. San Rafael Atlixco 186, CdMx 09340, Mexico.\\
$^7$Instituto de Ciencias del Espacio (ICE, CSIC), C. Can Magrans s.n., 08193 Cerdanyola del Vallès, Catalonia, Spain.\\
$^8$Institut d’Estudis Espacials de Catalunya (IEEC) C. Gran Capit\`a 2-4, Ed. Nexus, 08034 Barcelona, Spain. \\
$^9$Facultat de Física, Universitat de Barcelona, Martí i Franquès 1, 08028 Barcelona, Spain.%
}

\begin{abstract}

We use the two-flavor linear sigma model with quarks to study the phase structure of isospin asymmetric matter at zero temperature. The meson degrees of freedom provide the mean field chiral- and isospin-condensates on top of which we compute the effective potential accounting for constituent quark fluctuations at one-loop order. Using the renormalizability of the model, we absorb the ultraviolet divergences into suitable counter-terms that are added respecting the original structure of the theory. These counter-terms are determined from the stability conditions which require the effective potential to have minima in the condensates directions at the classical values, as well as the transition from the non-condensed to the condensed phase to be smooth as a function of the isospin chemical potential. We use the model to study the evolution of the condensates as well as the pressure, energy and isospin densities and the sound velocity as functions of the isospin chemical potential. The approach does a good average description up to isospin chemical potentials values not too large as compared to the vacuum pion mass.

\end{abstract}


\maketitle

\section{Introduction}\label{sec1}

Multiple implications of the remarkably rich phase structure of Quantum Chromodynamics (QCD) have been extensively explored over the last years. QCD at finite density is usually characterized by the baryon $\mu_B$ and the isospin $\mu_I$ chemical potentials. Nature provides us with physical systems at finite baryon densities with non zero $\mu_I$ in the form of isospin asymmetric matter, for example, compact astrophysical objects such as neutron stars. Because of this, along with the imminent arrival of new generation relativistic heavy-ion collision experiments at the FAIR~\cite{Agarwal:2022ydl} and NICA~\cite{MPD:2022qhn} facilities, the study of the phase structure in the temperature $T$ and the chemical potentials $\mu_B$ and $\mu_I$ has become an ideal subject of scrutiny within the heavy-ion and astroparticle physics communities~\cite{Fukushima:2010bq,Alford:2007xm}. 

A typical $T-\mu_B-\mu_I$ phase diagram is anticipated to be full of rich phase structures~\cite{NUPECC:2017lrp}. However, from the theoretical perspective, systems with finite $\mu_B$ are not straightforwardly accessible to the first-principle methods of Lattice QCD (LQCD), due to the well-known fermion determinant sign problem~\cite{Karsch:2001cy,Muroya:2003qs}. Hence, studies on the $\mu_B-\mu_I$ plane have been performed mainly using low energy effective models. These models have revealed the existence of an exciting phase structure that includes Gapless Pion Condensates (GPC), a Bose-Einstein Condensed (BEC) phase with gaped single particle excitations, a BEC-BCS crossover, etc~\cite{Son:2005qx,Mu:2010zz}. 
 
On the other hand, LQCD calculations for vanishing and even small $\mu_B$ do not suffer from the sign problem. These calculations have predicted the existence of a superfluid pion condensate phase for high enough $\mu_I$~\cite{Kogut:2002zg,Kogut:2002tm,Brandt:2016zdy,Brandt:2017zck,Brandt:2017oyy,Brandt:2018wkp}. At zero temperature, they show that a second order phase transition at a critical isospin chemical potential (corresponding to the vacuum pion mass), separates the hadron from the pion condensate phase~\cite{Brandt:2017oyy}. In addition to   LQCD, these phases are also found using chiral perturbation theory ($\chi$PT)~\cite{Son:2000xc,Son:2000by,Splittorff:2000mm,Loewe:2002tw,Loewe:2005yn,Fraga:2008be,Cohen:2015soa,Janssen:2015lda,Carignano:2016lxe,Lepori:2019vec,Adhikari:2020ufo,Adhikari:2020kdn,Adhikari:2020qda, ADHIKARI2020135352, PhysRevD.106.114017,Adhikari:2019mdk} , Hard Thermal Loop perturbation theory (HTLPt)~\cite{Andersen:2015eoa}, the Nambu-Jona-Lasinio (NJL) model~\cite{Frank:2003ve,Toublan:2003tt,Barducci:2004tt,He:2005sp,He:2005nk,He:2006tn,Ebert:2005cs,Ebert:2005wr,Sun:2007fc,Andersen:2007qv,Abuki:2008wm,Mu:2010zz,Xia:2013caa,Khunjua:2018jmn,Khunjua:2018sro,Khunjua:2017khh,Ebert:2016hkd} and its Polyakov loop (PNJL) extended version~\cite{Mukherjee:2006hq,Bhattacharyya:2012up}, the quark meson model (QMM)~\cite{Kamikado:2012bt,Ueda:2013sia,Stiele:2013pma,Adhikari:2018cea} and other low energy effective models exploiting functional RG studies~\cite{Braun:2022olp}. Calculations using a LQCD equation of state for finite $\mu_I$ have investigated the viability of the existence of pion stars, with a pion condensate as the dominant core constituent~\cite{Carignano:2016lxe,Brandt:2018bwq}. Since LQCD calculations with $\mu_I \neq 0,~\mu_B=\mu_s=T=0$ can be carried out without being hindered by the sign problem, they can be used as a benchmark to test effective model predictions. For example, recently, the NJL model has been used in this domain and it has been found that results agree exceptionally well with LQCD results~\cite{Avancini:2019ego,Lopes:2021tro}. 

In this work we study  another effective QCD model, the Linear Sigma Model with quarks (LSMq), extended to consider a finite $\mu_I$ to describe the properties of strongly interacting systems with an isospin imbalance. The LSMq is a renormalizable theory that explicitly implements the QCD chiral symmetry. It has been successfully employed to study the chiral phase transition at finite $T$ and $\mu_B$~\cite{Ayala:2021tkm,Ayala:2019skg,Ayala:2017ucc,Ayala:2014jla}, as well as in the presence of a magnetic field~ { \cite{Ayala:2021nhx,Ayala:2020dxs,Ayala:2020muk,Ayala:2018zat,Ayala:2015lta,Ayala:2014gwa,Ayala:2014iba,Ayala:2014mla,Das:2019ehv,Ghosh:2022xtv}}.
The Linear Sigma Model has been used at finite $\mu_I$, albeit considering the meson degrees of freedom as an effective classical background, in the Hartree or Hartree Fock approximations within the Cornwall-Jackiw-Tomboulis (CJT) formalism~\cite{Mao:2013gva}. In contrast, in the LSMq mesons are treated as dynamical fields able to contribute to quantum fluctuations. Part of the reason for other models to avoid considering mesons as dynamical fields, for example the QMM, is that when mesons become true quantum fields and chiral symmetry is only spontaneously broken, their masses are subject to change as a result of medium effects.
During this change, the meson square masses can become zero or even negative. At zero temperature, this drawback is avoided by considering an explicit symmetry breaking term that provides pions with a vacuum finite mass. At finite temperature, the plasma screening effects need to also be included. 

In this work we use the LSMq to describe the evolution of the chiral and isospin (pion) condensates, as well as thermodynamical quantities such as pressure, isospin and energy densities and the sound velocity at zero temperature and finite $\mu_I$. We restrict ourselves to considering only the effects of fermion quantum fluctuations, reserving for a future work the inclusion of meson quantum fluctuations effects. We make use of the renormalizability of the LSMq and describe in detail the renormalization procedure which is achieved by implementing the stability conditions. The results thus obtained are valid for the case where $\mu_I^2$ is small compared to the sum of the squares of the chiral and isospin condensates multiplied by the square of the boson-fermion coupling constant $g$.

The work is organized as follows: In Sec.~\ref{secII} we write the LSMq Lagrangian using degrees of freedom appropriate to describe an isospin imbalanced system. We work with an explicit breaking of the chiral symmetry introducing a vacuum pion mass and expanding the charged pion fields around the values of their condensates. The effective potential is constructed by adding to the tree-level potential the one-loop contribution from the fermion degrees of freedom. Renormalization is carried out by introducing counter-terms to enforce that the tree-level structure of the effective potential is preserved by loop corrections. We first work out explicitly the treatment in the condensed phase to then work out the non-condensed phase. In Sec.~\ref{secIII} we study the condensates evolution with $\mu_I$ as well as that of the pressure, isospin and energy density and the sound velocity, and compare to recent LQCD results. We finally summarize and conclude in Sec.~\ref{secIV}. We reserve for a follow up work the computation of the meson quantum fluctuations as well as finite temperature effects. The appendix is devoted to the explicit computation of the one-loop fermion contribution to the effective potential.
\section{ LSMq at finite isospin chemical potential}\label{secII}

The LSMq is an effective theory that captures the approximate chiral symmetry of QCD. It describes the interactions among small-mass mesons and constituent quarks. We start with a Lagrangian invariant under $SU(2)_{L}\times SU(2)_{R}$ chiral transformations
\begin{eqnarray}
\mathcal{L}&=&\frac{1}{2}(\partial_{\mu}\sigma)^{2}+\frac{1}{2}(\partial_{\mu}\vec{\pi})^{2}+\frac{a^{2}}{2}(\sigma^{2}+\vec{\pi}^{2})-\frac{\lambda}{4}(\sigma^{2}
+\vec{\pi}^{2})^{2}\nonumber\\
&+&i\bar{\psi}\gamma^{\mu}\partial_{\mu}\psi-ig\bar{\psi}\gamma^{5}\vec{\tau}\cdot \vec{\pi}\psi -g\bar{\psi}\psi\sigma,
\label{lsmqlag1}
\end{eqnarray}
where  
$\vec{\tau}=(\tau_{1},\tau_{2},\tau_{3})$
are the Pauli matrices, \begin{eqnarray}
\psi_{L,R}= \begin{pmatrix} u \\ d \end{pmatrix}_{L,R},
\label{doublet}
\end{eqnarray}is a  $SU(2)_{L,R}$ doublet, $\sigma$ is a real scalar field and $\vec{\pi}=(\pi_{1},\pi_{2},\pi_{3})$ is a triplet of real scalar fields. $\pi_3$ corresponds to the neutral pion whereas the charged ones are represented by the combinations
\begin{eqnarray}
    \pi_{-}=\frac{1}{\sqrt{2}}(\pi_{1}+i\pi_{2}), \quad \pi_{+}=\frac{1}{\sqrt{2}}(\pi_{1}-i\pi_{2}).
    \label{pi1pi2basis}
\end{eqnarray}
The parameters $a^2$, $\lambda$ and $g$ are real and positive definite. Equation~(\ref{lsmqlag1}) can be written in terms of the charged and neutral-pion degrees of freedom as
\begin{eqnarray}
\nonumber
\mathcal{L}&=&\frac{1}{2}[(\partial_{\mu}\sigma)^{2}+(\partial_{\mu}\pi_{0})^{2}] +\partial_{\mu}\pi_{-}\partial^{\mu}\pi_{+}+\frac{a^{2}}{2}(\sigma^{2}+\pi_{0}^{2})\nonumber\\
&+&a^{2}\pi_{-}\pi_{+} -\frac{\lambda}{4}(\sigma^{4}+4\sigma^{2}\pi_{-}\pi_{+}+2\sigma^{2}\pi_{0}^2+4\pi_{-}^{2}\pi_{+}^{2}\nonumber\\
&+&4\pi_{-}\pi_{+}\pi_{0}^{2}+\pi_{0}^{4})+i\bar{\psi}\slashed{\partial}\psi-g\bar{\psi}\psi\sigma-ig\bar{\psi}\gamma^{5}(\tau_{+}\pi_{+}\nonumber\\
&+&\tau_{-}\pi_{-}+\tau_{3}\pi_{0})\psi,
\label{lsmqlag2}    
\end{eqnarray}
where we introduced the combination of Pauli matrices
\begin{eqnarray}
    \tau_{+}=\frac{1}{\sqrt{2}}(\tau_{1}+i\tau_{2}), \quad \tau_{-}=\frac{1}{\sqrt{2}}(\tau_{1}-i\tau_{2}).
\end{eqnarray}
The Lagrangian in Eq.~(\ref{lsmqlag2}) possesses the following symmetries: A $SU(N_c)$ global color symmetry, a $SU(2)_{L}\times SU(2)_{R}$ chiral symmetry and a $U(1)_B$ symmetry. The sub-index of the latter  emphasizes that the conserved charge is the baryon number $B$. A conserved isospin charge can be added to the LSMq Hamiltonian, multiplied by the isospin chemical potential $\mu_I$. The result is that the Lagrangian gets modified such that the ordinary derivative becomes a covariant derivative~\cite{Mannarelli:2019hgn}
\begin{equation}
    \partial_{\mu} \to D_{\mu}= \partial_{\mu}+i\mu_{I} \delta_{\mu}^{0}, \quad  \partial^{\mu} \to D^{\mu}= \partial^{\mu}-i\mu_{I} \delta_{0}^{\mu},
\end{equation}
As a result, Eq.~(\ref{lsmqlag2}) is modified to read as
\begin{eqnarray}
\mathcal{L}&=&\frac{1}{2}[(\partial_{\mu}\sigma)^{2}+(\partial_{\mu}\pi_{0})^{2}]+D_{\mu}\pi_{-}D^{\mu}\pi_{+}+\frac{a^{2}}{2}(\sigma^{2}+\pi_{0}^{2})\nonumber\\
&+&a^{2}\pi_{-}\pi_{+}-\frac{\lambda}{4}\left(\sigma^{4}+4\sigma^{2}\pi_{-}\pi_{+}+2\sigma^{2}\pi_{0}^2+4\pi_{-}^{2}\pi_{+}^{2}\right.\nonumber\\
&+&\left. 4\pi_{-}\pi_{+}\pi_{0}^{2}+\pi_{0}^{4}\right)+i\bar{\psi}\slashed{\partial}\psi-g\bar{\psi}\psi\sigma+ \bar{\psi}\mu_I\tau_3\gamma_0\psi\nonumber\\
&-&ig\bar{\psi}\gamma^{5}(\tau_{+}\pi_{+}+\tau_{-}\pi_{-}+\tau_{3}\pi_{0})\psi .
\label{lsmqlagwmui}    
\end{eqnarray}

Because of the spontaneous breaking of the chiral symmetry in the Lagrangian given in Eq.~(\ref{lsmqlagwmui}), the $\sigma$ field acquires a non-vanishing vacuum expectation value
\begin{equation*}
   \sigma \rightarrow \sigma+v.
\end{equation*}
To make better contact with the meson vacuum properties and to include a finite vacuum pion mass, $m_0$, we can add an explicit symmetry breaking term in the Lagrangian such that
\begin{equation}
    \mathcal{L} \to \mathcal{L'}=\mathcal{L} +h(\sigma + v).
    \label{lsmqlagwesb}
\end{equation}
The constant $h$ is fixed by requiring that the model expression for the neutral vacuum pion mass squared in the non-condensed phase, Eq.~(\ref{massesferand neu}b), corresponds to $m_0^2$. Recall that in the non-condensed phase, the tree-level potential is
\begin{equation*}
V_{\text{tree}}=–\frac{a^2}{2} v^2 + \frac{\lambda}{4} v^4  - h v.
\end{equation*}
The condensate $v_0$ is obtained from
\begin{equation*}
\frac{dV_{\text{tree}}}{dv} = (\lambda v^3 – a^2 v -h)_{v=v_0}=0,
\end{equation*}
or
\begin{equation*}
v_0(\lambda v_0^2 – a^2)=h
\end{equation*}
The quantity in between parenthesis, according to Eq.~(11b), is precisely the square of the vacuum pion mass, $m_0^2$. Therefore
\begin{equation}
h=m_0^2 v_0
\nonumber
\end{equation}
Also, notice that 
\begin{equation}
a^2 + m_0^2 = a^2 +\lambda v_0^2 – a^2 = \lambda v_0^2,
\nonumber
\end{equation}
or
\begin{equation}
v_0= \sqrt{\frac{a^2+m_0^2}{\lambda}}.
\nonumber
\end{equation}
This yields
\begin{eqnarray}
h&=&m_0^2\sqrt{\frac{a^2+ m_0^2}{\lambda}},\nonumber\\
&\equiv&m_0^2f_\pi,
\label{expressionforh}
\end{eqnarray}
where in the second line we have used the Partially Conserved Axial Current (PCAC) statement to identify $m_0^2f_\pi$ with the small symmetry breaking term represented in the LSMq by $h$, with $f_\pi$ the pion decay constant. Equation~(\ref{expressionforh}) provides a relation for the model parameters $a$ and $\lambda$ in terms of $f_\pi$.

Before diving into the formalism details, here we first pause to discuss the symmetry properties of the theory. Notice that the introduction of $\mu_I$ and $h$ modifies the structure of the effective Lagrangian given in Eq.~(\ref{lsmqlagwesb}). In the presence of a finite $\mu_I$, the $U(1)_B\times SU(2)_{L} \times SU(2)_{R}$ symmetry is reduced to $U(1)_B\times U(1)_{I_3L}\times U(1)_{I_3R}$ for $h=0$, and to $U(1)_B\times U(1)_{I_3}$ for $h\neq 0$, thereby representing the explicit breaking of the chiral symmetry~\cite{RevModPhys.64.649}. The notation also emphasizes that the third component of the isospin charge, $I_3$, corresponds to the generator of the remaining symmetry $U(1)_{I_3}$. Since in the present work, we are interested in the dynamics of the pion fields, further simplifications in the pseudoscalar channels can be obtained using the ansatz $\langle \bar\psi i\gamma_5\tau_3\psi\rangle=0$ combined with $\langle \bar{u}i\gamma_5 d\rangle = \langle \bar{d}i\gamma_5 u\rangle^* \neq 0$~\cite{Mu:2010zz}. This further breaks the residual $U(1)_{I_3}$ symmetry and corresponds to a Bose-Einstein condensation of the charged pions. Then, the charged pion fields can be referred from their condensates as 
\begin{equation}
\pi_{+}\to \pi_{+}+\frac{\Delta}{\sqrt{2}} e^{i\theta}, \quad \pi_{-}\to \pi_{-}+\frac{\Delta}{\sqrt{2}} e^{-i\theta},
\end{equation}
where the phase factor $\theta$ indicates the direction of the $U(1)_{I_3}$ symmetry breaking. We take $\theta=\pi$ for definitiveness. The shift in the sigma field produces that the fermions and neutral bosons acquire masses given by
\begin{subequations}
    \begin{eqnarray}
m_f&=&gv\\
m_{\pi^0}^2&=&\lambda v^2 -a^2 +\lambda\Delta^2\\
m_{\sigma}^2&=&3\lambda v^2 -a^2 +\lambda\Delta^2.
\end{eqnarray}
\label{massesferand neu}
\end{subequations}
The charged pions also acquire masses. However, in the condensed phase ($\Delta\neq 0$) they need to be described in terms of the $\pi_{1, 2}$ fields~\cite{Schmitt:2014eka}. Since for our purposes, pions are not treated as quantum fluctuations, hereby we just notice that, as a consequence of the breaking of the $U(1)_{I_3}$ symmetry, one of these fields becomes a Goldstone boson. In the absence of the explicit symmetry breaking term in the Lagrangian of Eq.~(\ref{lsmqlagwesb}), this mode's mass would vanish. However, a finite $h$ prevents this mode from being massless.

\subsection{Condensed phase}

In the condensed phase the tree-level potential, extracted from Eqs.~(\ref{lsmqlagwmui}) and~(\ref{lsmqlagwesb}), can be written as 
\begin{equation}
    V_{\rm tree}=-\frac{a^2}{2}\left(v^{2}+\Delta^2\right)+\frac{\lambda}{4}\left(v^2+\Delta^2 \right)^2-\frac{1}{2}\mu_I^2\Delta^2-hv.
\label{treeeffectivepotential}
\end{equation}

The fermion contribution to the one-loop effective potential becomes
\begin{equation}
\sum_{f=u,d}V_{f}^1= -2N_c\int\frac{d^3k}{(2\pi)^3}\left[E_\Delta^u +E_\Delta^d\right],
\label{fromfermions}
\end{equation}
with (see Appendix A)
\begin{subequations}
\begin{eqnarray}
E_\Delta^u &=& \left\{\left(\sqrt{k^2+m_f^2}+\mu_I\right)^2+g^2\Delta^2\right\}^{1/2}, \\
E_\Delta^d &=& \left\{\left(\sqrt{k^2+m_f^2}-\mu_I\right)^2+g^2\Delta^2\right\}^{1/2},
\end{eqnarray}
\label{fermionenergies}
\end{subequations}
where we chose that
\begin{eqnarray}
\mu_d&=&\mu_I\nonumber\\
\mu_u&=&-\mu_I.
\label{choice}
\end{eqnarray}
Equation~(\ref{fromfermions}) is ultraviolet divergent. Ultraviolet divergences are a common feature of loop vacuum contributions. However, since Eq.~(\ref{fromfermions}) depends on $\mu_I$, this divergence needs to be carefully treated given that matter contributions cannot contain ultraviolet divergences. To identify the divergent terms, we work in the approximation whereby the fermion energies, Eqs.~(\ref{fermionenergies}), are expanded in powers of $\mu_I^2/[g^2(v^2+\Delta^2)]$. Considering terms up to $\mathcal{O}(\mu_I^4)$, we obtain 
\begin{equation}
\begin{split}
\sum_{f=u,d} E_\Delta^f&\simeq 2\sqrt{k^2+m_f^2+g^2\Delta^2}+\frac{\mu_I^2g^2\Delta^2}{(k^2+m_f^2+g^2 \Delta^2)^{3/2}} \\
&+\frac{\mu_I^4 \left[4(k^2+m_f^2)g^2 \Delta^2-g^4\Delta^4 \right]}{4 \left(k^2+m_f^2+g^2\Delta^2 \right)^{7/2}}   +\mathcal{O}(\mu_I^{6}).
\end{split}
\label{expaninmu}
\end{equation}
Notice that the ultraviolet divergent part corresponds only to the first and second terms on the right-hand side of Eq.~(\ref{expaninmu}). In this approximation, and up to terms of order $\mu_I^2$, the expression for the leading fermion contribution to the one-loop effective potential is given by
\begin{eqnarray}
\sum_{f=u,d}V_{f}^1&=& -2N_c\int\frac{d^3k}{(2\pi)^3}\Big( 2\sqrt{k^2+m_f^2+g^2\Delta^2}\nonumber\\
&+&\frac{\mu_I^2g^2\Delta^2}{(k^2+m_f^2+g^2 \Delta^2)^{3/2}}\Big)
\label{V1finf}
\end{eqnarray}

This expression can be readily computed using dimensional regularization in the $\overline{\mbox{MS}}$ scheme, with the result (see Appendix A)
\begin{eqnarray}
\sum_{f=u,d}V_{f}^1&=&2N_c\frac{g^4\left(v^2+\Delta^2\right)^2}{(4\pi)^2}\left[\frac{1}{\epsilon}+\frac{3}{2}+\ln\left(\frac{\Lambda^2/g^2}{v^2+\Delta^2}\right)\right]\nonumber\\
&-&2N_c\frac{g^2\mu_I^2\Delta^2}{(4\pi)^2}\left[\frac{1}{\epsilon}+\ln\left(\frac{\Lambda^2/g^2}{v^2+\Delta^2}\right)\right],
\label{explVdiv}
\end{eqnarray}
where $N_c=3$ is the number of colors, $\Lambda$ is the dimensional regularization ultraviolet scale and the limit $\epsilon\to 0$ is to be understood. The explicit computation of Eq.~(\ref{explVdiv})  is described also in Appendix A. Notice that Eq.~(\ref{explVdiv}) contains an ultraviolet divergence proportional to $\mu_I^2\Delta^2$. Since a term with this same structure is already present in the tree-level potential, Eq.~(\ref{treeeffectivepotential}), it is not surprising that this ultraviolet divergence can be handled by the renormalization procedure with the introduction of a counter-term with the same structure, as we proceed to show.

To carry out the renormalization of the effective potential up to one-loop order, we introduce counter-terms that respect the structure of the tree-level potential and determine them by accounting for the stability conditions. The latter are a set of conditions satisfied by the tree-level potential and that must be preserved when considering loop corrections. These conditions require that the position of the minimum
in the $v$- and $\Delta$-directions remain the same as the tree-level potential ones. Notice that this approach is different from the one followed in Ref.~\cite{PhysRevD.78.116008}, where the counter-terms are determined by requiring the finiteness of the propagator and the four-point boson vertex. As a result a shift of the onset of pion condensation happens when the coupled
equations that determine the condensates receive loop corrections.

The tree-level minimum in the $v$, $\Delta$ plane is found from
\begin{subequations}
    \begin{eqnarray}
\!\!\!\!\!\!\!\!\left.\frac{\partial V_{\rm tree}}{\partial v}=\left[\lambda v^3 -(a^2 - \lambda\Delta^2)v-h\right]\right|_{v_0,\,\Delta_0}&=&0\\
\!\!\!\!\!\!\!\!\left.\frac{\partial V_{\rm tree}}{\partial \Delta}=\left[\lambda\Delta^2-(\mu_I^2-\lambda v^2 + a^2)\right]\right|_{v_0,\,\Delta_0}&=&0.
\end{eqnarray}
\label{minima}
\end{subequations}
Notice that the second of Eqs.~(\ref{minima}) admits a real, non-vanishing solution, only when 
\begin{eqnarray}
\mu_I^2 > \lambda v^2 - a^2 = m_{0}^2,
\label{conditionmu}
\end{eqnarray}
which means that a non-zero isospin condensate is developed only when, for positive values of the isospin chemical potential, the latter is larger than the vacuum pion mass. This is what we identify as the condensed phase. The simultaneous solutions of Eqs.~(\ref{minima}) are
\begin{subequations}
   \begin{eqnarray}
v_0&=&\frac{h}{\mu_I^2},\\
\Delta_0&=&\sqrt{\frac{\mu_I^2}{\lambda} - \frac{h^2}{\mu_I^4} + \frac{a^2}{\lambda}}.
\end{eqnarray} 
\label{simultaneous}
\end{subequations}
Hereafter, we refer to the expressions in Eq.~(\ref{simultaneous}) as the classical solution.

The effective potential, up to one-loop order in the fermion fluctuations, including the counter-terms, can be written as
\begin{eqnarray}
V_{\rm eff}&=&V_{\rm tree}+\sum_{f=u,d}V_{f}^1-\frac{\delta\lambda}{4}(v^2+\Delta^2)^2\nonumber\\
&+&\frac{\delta a}{2}(v^2+\Delta^2)+\frac{\delta}{2}\Delta^2\mu_I^2.
\label{withcounterterms}
\end{eqnarray}
The counter-terms $\delta\lambda$ and $\delta$ are determined from the {\it gap equations} 
\begin{subequations}
    \begin{eqnarray}
\left.\frac{\partial V_{\rm eff}}{\partial v}\right|_{v_0,\,\Delta_0}&=&0,\\
\left.\frac{\partial V_{\rm eff}}{\partial \Delta}\right|_{v_0,\,\Delta_0}&=&0.
\end{eqnarray}
\label{stabilitycond}
\end{subequations}
These conditions suffice to absorb the infinities of Eq.~(\ref{explVdiv}). The counter-term $\delta a$ is determined by requiring that the slope of $V_{\rm eff}$ vanishes at $\mu_I=m_0$, 
\begin{eqnarray}
\left.\frac{\partial V_{\rm eff}}{\partial \mu_I}\right|_{\mu_I=m_0}=0,
\label{thirdcond}
\end{eqnarray}
or in other words, that the transition from the non-condensed to the condensed phase be smooth. The resulting effective potential is also $\Lambda$-independent. This can be seen by noticing that the coefficients of the $1/\epsilon$ terms are common to those of the $\ln(\Lambda^2)$ terms. Since the counter-terms cancel the $1/\epsilon$ divergence, they also  cancel the $\ln(\Lambda^2)$ dependence.

\subsection{Non-condensed phase}

In the non-condensed phase, $0\leq \mu_I\leq m_{0}$, the only allowed solution for the second of Eqs.~(\ref{minima}) is $\Delta =0$. For this case, the first of Eqs.~(\ref{minima}) becomes a cubic equation in $v$. The only real solution is
\begin{eqnarray}
\tilde{v}_0&=&\frac{(\sqrt{3} \sqrt{27 h^2 \lambda ^4-4 a^6 \lambda ^3}+9 h \lambda ^2)^{1/3}}{(18)^{2/3} \lambda}\nonumber\\
&+&\frac{(2/3)^{1/3} a^2}{(\sqrt{3} \sqrt{27 h^2 \lambda ^4-4 a^6 \lambda ^3}+9 h \lambda ^2)^{1/3}}.
\label{realv0}
\end{eqnarray}
In the limit when $h$ is taken as small one gets
\begin{eqnarray}
\tilde{v}_0\simeq\frac{a}{\sqrt{\lambda}}+\frac{h}{2a^2},
\label{smallh}
\end{eqnarray}
an approximation that some times is considered. However, hereafter we work instead with the full expression given by Eq.~(\ref{realv0}). 

The effective potential $V^{\rm noncond}_{\rm eff}$ up to one-loop order can be obtained from the corresponding one in the condensed phase, by setting $\Delta=0$. 
Therefore, we can write
\begin{eqnarray}
V^{\rm noncond}_{\rm eff}&=&\frac{\lambda}{4}v^4-\frac{a^2}{2}v^2-hv-\frac{\tilde{\delta}_1}{4}v^4
+\frac{\tilde{\delta}_2}{2}v^2\nonumber\\
&+&
2N_c\frac{g^4v^4}{(4\pi)^2}\left[\frac{1}{\epsilon}+\frac{3}{2}+\ln\left(\frac{\Lambda^2}{g^2v^2}\right)\right].
\label{Vfullnocond}
\end{eqnarray}
In this case, only two conditions are needed to stabilize the vacuum. We take these as the requirement that the position and curvature of $V^{\rm noncond}_{\rm eff}$ remain at its classical value when evaluated at $\tilde{v}_0$, namely,
\begin{subequations}
    \begin{eqnarray}
\left.\frac{\partial V^{\rm noncond}_{\rm eff}}{\partial v}\right|_{\tilde{v}_0}&=&0\\
\left.\frac{\partial^2 V^{\rm noncond}_{\rm eff}}{\partial v^2}\right|_{\tilde{v}_0}&=&3\lambda \tilde{v}_0^2-a^2,
\end{eqnarray}
\label{stabilitynoncond}
\end{subequations}
from where the counter-terms $\tilde{\delta}_1$, $\tilde{\delta}_2$ can be determined. Therefore, in the non-condensed phase, in addition to $\Delta=0$, the $v$-condensate is simply given by the constant $\tilde{v}_0$ given in Eq.~(\ref{realv0}). As for the case of the condensed phase, in the non-condensed phase the effective potential is ultraviolet finite as well as $\Lambda$-independent.


\section{Thermodynamics of the condensed phase}\label{secIII}

\begin{figure}[t]
    \begin{center}
    \includegraphics[scale=0.85]{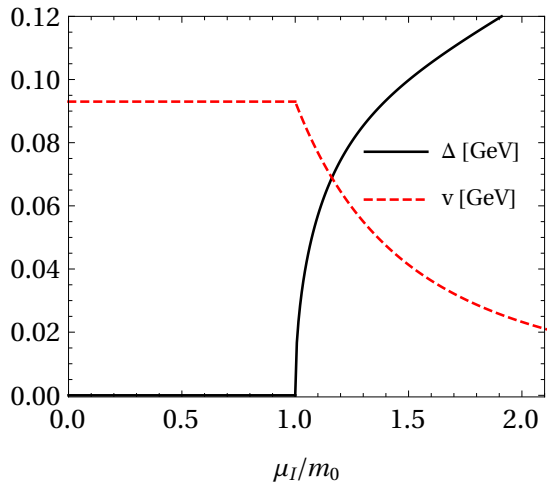}
    \end{center}
    \caption{$v$- and $\Delta$-condensates as functions of the scaled variable $\mu_I/m_0$. For $\mu_I\geq m_0$, the $v$-condensate decreases while the $\Delta$-condensate increases.}
    \label{fig1}
\end{figure}

Armed with the expressions for the effective potential, we can now proceed to study the dependence of the condensates as well as of the thermodynamical quantities as functions of $\mu_I$. Since the $\mu_I$-dependence in the non-condensed phase is trivial, we concentrate in the description of the behavior of these quantities in the condensed phase.

The model requires fixing three independent parameters: the boson self-coupling $\lambda$, the boson-fermion coupling $g$ and the mass parameter $a$. 
For this purpose, notice that from Eqs.~(\ref{massesferand neu}) in vacuum we have
\begin{equation*}
m_\sigma^2 – 3m_0^2 = 3\lambda v_0^2 – a^2 – 3\lambda v_0^2 + 3a^2                                            = 2a^2,
\nonumber
\end{equation*}
or
\begin{equation*}
a=\sqrt{\frac{m_\sigma^2 – 3m_0^2}{2}}.
\nonumber
\end{equation*}
Also 
\begin{equation*}
g = \frac{m_q}{v_0} = \frac{m_q}{f_\pi},
\nonumber
\end{equation*}
and $\lambda$ is obtained from Eq.~(\ref{expressionforh}) as
\begin{equation*}
\lambda=\frac{m_\sigma^2-m_0^2}{2f_\pi^2}
\nonumber
\end{equation*}
For $m_q=235$ MeV, $m_\sigma= 400$ MeV, $m_0=140$ MeV and $f_\pi=93$ MeV, one readily obtains $\lambda=8.12$, $a=225$ MeV and $g=2.53$.
The phase space for these parameters is limited since for certain combinations, the gap equation conditions in the $v$-$\Delta$ plane become saddle points rather than global minima. A more exhaustive search in the parameter space to optimize the parameter choice will be discussed elsewhere.

\begin{figure}[t]
    \begin{center}
    \includegraphics[scale=0.85]{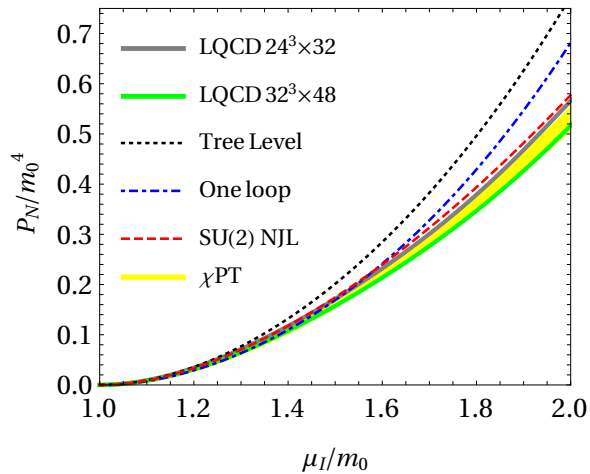}
    \end{center}
    \caption{Normalized pressure as a function of the scaled variable $\mu_I/m_0$. Shown are the tree-level and one-loop fermion improved pressures compared to the results from Refs.~\cite{Avancini:2019ego,Adhikari:2019mdk} together with the LQCD results from a private communication with the authors of Refs.~\cite{Brandt:2022fij,Brandt:2022hwy}.}
    \label{fig2}
\end{figure}
\begin{figure}[t]
    \begin{center}
    \includegraphics[scale=0.4]{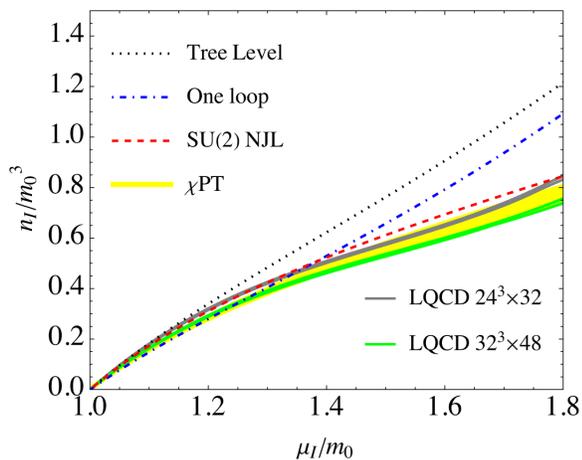}
    \end{center}
    \caption{Normalized isospin density as a function of the scaled variable $\mu_I/m_0$. Shown are the tree-level and one-loop fermion improved effective potentials compared to a recent $SU(2)$ NJL calculation~\cite{Avancini:2019ego}, two-flavor $\chi$PT~\cite{Adhikari:2019mdk} and the LQCD results from Ref.~\cite{Brandt:2022fij,Brandt:2022hwy}.}
    \label{fig3}
\end{figure}
\begin{figure}[b]
    \begin{center}
    \includegraphics[scale=0.85]{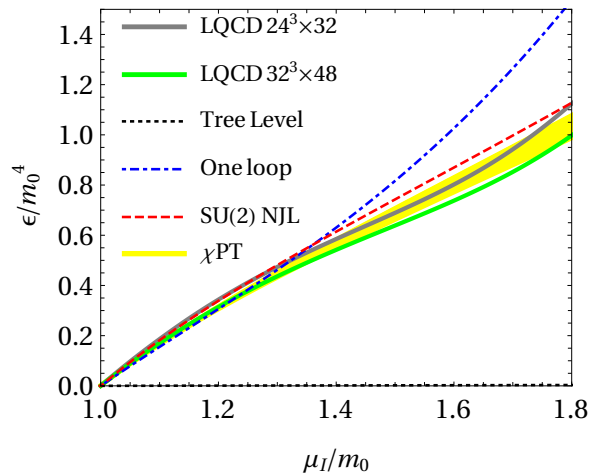}
    \end{center}
    \caption{Normalized energy density as a function of the scaled variable $\mu_I/m_0$. Shown are the tree-level and one-loop fermion improved effective potentials compared to the results from Refs.~\cite{Avancini:2019ego,Adhikari:2019mdk} together with the LQCD results from Ref.~\cite{Brandt:2022fij,Brandt:2022hwy}.}
    \label{fig4}
\end{figure}
\begin{figure}[t]
    \begin{center}
    \includegraphics[scale=0.4]{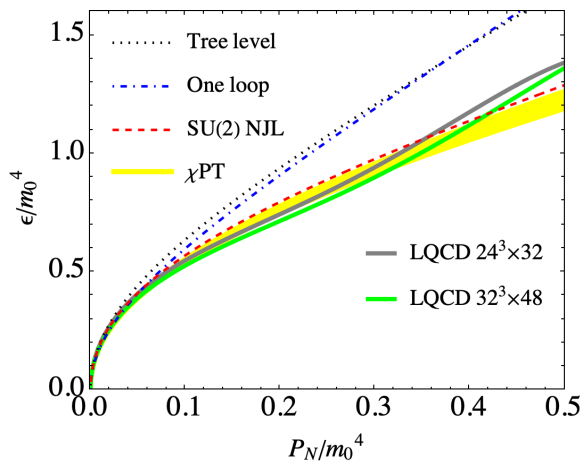}
    \end{center}
    \caption{Equation of state, pressure vs. energy density. Shown are the tree-level and one-loop fermion improved effective potentials compared to the results from Refs.~\cite{Avancini:2019ego,Adhikari:2019mdk} together with the LQCD results from Ref.~\cite{Brandt:2022fij,Brandt:2022hwy}. 
    }
    \label{fig5}
\end{figure}

Figure~\ref{fig1} shows the $v$- and $\Delta$-condensates as functions of the scaled variable $\mu_I/m_0$. The behavior is qualitatively as expected: for $\mu_I\geq m_0$, the $v$-condensate decreases while the $\Delta$-condensate increases.

Figure~\ref{fig2} shows the normalized pressure, defined as the negative of the effective potential referred from its value at $\mu_I=m_0$, as a function of the scaled variable $\mu_I/m_0$ and divided by $m_0^4$. Shown are the results obtained by using the tree-level and the fermion one-loop corrected effective potentials, compared to the results from a $SU(2)$ NJL model~\cite{Avancini:2019ego}, a $SU(2)$ $\chi$PT~\cite{Adhikari:2019mdk}, and the LQCD results from Refs.~\cite{Brandt:2022fij,Brandt:2022hwy}. The $\chi$PT results consist of a leading order (LO) and a next to leading order (NLO) calculations. The NLO ones depend on low energy constants. We show both results with a yellow band~\cite{Adhikari:2019mdk}. The LQCD results from Refs.~\cite{Brandt:2022fij,Brandt:2022hwy} were obtained using simulations with three dynamical flavors at physical quark masses. The labels $24^3\times 32$ and $32^3\times 48$ refer to the lattice sizes in the space and time directions. The lattice spacing of the former is $a \approx 0.22$ fm and for the latter $a \approx 0.15$ fm. The pion mass used in the LQCD calculation is $m_0=135$ MeV. However, notice that since we report the results as functions of the scaled variable $\mu_I/m_0$, these can be safely compared to the LQCD results. Notice that the one-loop improved calculation does a better description than the tree-level one and that deviations from the LQCD result appear for $\mu_I\gtrsim 1.5\ m_0$.

Figure~\ref{fig3} shows the normalized isospin density, $n_I=dP/d\mu_I$, divided by $m_0^3$ as a function of the scaled variable $\mu_I/m_0$ compared to results obtained using the tree-level potential as well as to the results from Ref.~\cite{Avancini:2019ego} together with the LQCD results from Refs.~\cite{Brandt:2022fij,Brandt:2022hwy}. Notice that the one-loop improved calculation is close to the NJL one up to $\mu_I\sim 1.5\ m_0$ but the latter does a better job describing the LQCD results for $\mu_I\gtrsim 1.5\ m_0$. However, it is fair to say that none of the current calculations reproduce the change of curvature that seems to be present in the LQCD result.

Figure~\ref{fig4} shows the normalized energy density, $\epsilon/m_0^4$, with $\epsilon$ defined as
\begin{equation*}
\epsilon = -P + n_I\mu_I,
\end{equation*}
as a function of the scaled variable $\mu_I/m_0$, compared to the results from Ref.~\cite{Avancini:2019ego} together with the LQCD results from Refs.~\cite{Brandt:2022fij,Brandt:2022hwy}. Although the change in curvature shown by the LQCD results is not described by the present calculation, it is fair to say that neither the NJL calculation captures such trend. The one-loop improved calculation does a better average description of the LQCD result although deviations appear for $\mu_I\gtrsim 1.5\ m_0$.

Figure~\ref{fig5} shows the equation of state, pressure vs. energy density, compared to the results from Ref.~\cite{Avancini:2019ego} together with the LQCD results from Refs.~\cite{Brandt:2022fij,Brandt:2022hwy}. Notice that for the latter, the vacuum pion mass is taken as $m_0=135$ MeV. As can be seen, the initial increasing trend of LQCD results is properly described by the low-energy models considered. Given that the accuracy of our results is limited to the low $\mu_I$ domain the NJL calculation does a better description of the LQCD results.

\begin{figure}[t]
    \begin{center}
    \includegraphics[scale=0.4]{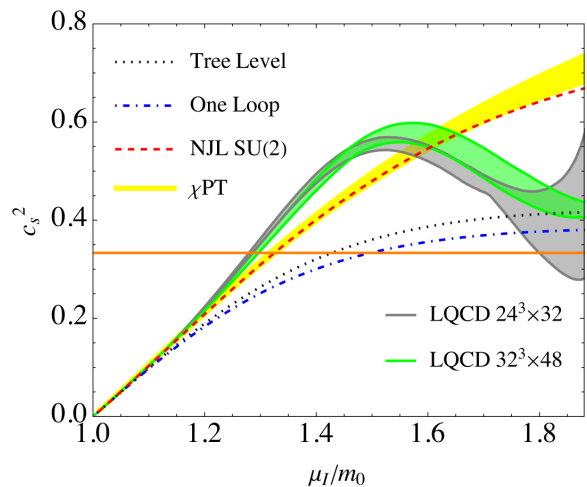}
    \end{center}
    \caption{Square of the speed of sound as a function of the scaled variable $\mu_I/m_0$. Shown are the tree-level and one-loop fermion improved effective potentials compared to recent $SU(2)$ NJL, $\chi$PT and the LQCD results from Ref.~\cite{Brandt:2022fij,Brandt:2022hwy}. The conformal bound is shown as a horizontal line.}
    \label{fig6}
\end{figure}

Figure~\ref{fig6} shows the square of the speed of sound, $c_s^2$, as a function of the scaled variable $\mu_I/m_0$. Shown are the one-loop results compared to the results from Ref.~\cite{Avancini:2019ego} together with the LQCD results from Refs.~\cite{Brandt:2022fij,Brandt:2022hwy}. The apparent peak in the LQCD results is not reproduced by any model. However, notice that for the range of shown $\mu_I$ values, the one-loop improved result is above, although closer to the conformal bound, shown as a horizontal line at $c_s^2=1/3$.

\begin{figure}[t]
    \begin{center}
    \includegraphics[scale=0.4]{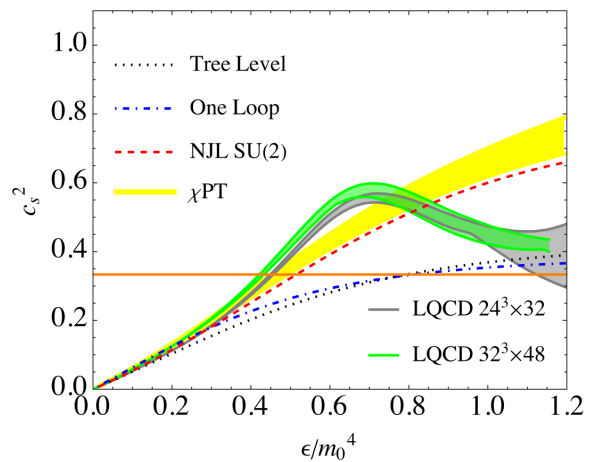}
    \end{center}
    \caption{Square of the speed of sound as a function of the scaled energy density. Shown are the tree-level and one-loop fermion improved effective potentials compared to recent $SU(2)$ NJL, $\chi$PT and the LQCD results from Ref.~\cite{Brandt:2022fij,Brandt:2022hwy}. The conformal bound is shown as a horizontal line.}
    \label{fig7}
\end{figure}

Figure~\ref{fig7} also shows $c_s^2$, this time as a function of  $\epsilon/m_0^4$ compared with results from Ref.~\cite{Avancini:2019ego} together with the LQCD results from Refs.~\cite{Brandt:2022fij,Brandt:2022hwy}. Although for lower values of the energy density the tree-level line lies below all other curves, after crossing the conformal bound, the one-loop improved result remains closer to the latter.


\section{Summary and Conclusions}\label{secIV}

In this work we have used the LSMq, with two constituent quark flavors, to study the phase structure of isospin asymmetric matter at zero temperature. The meson degrees of freedom are taken as providing the mean field on top of which we include quantum constituent quark fluctuations at one-loop order. We have used the renormalization of the LSMq to absorb the ultraviolet divergences with the addition of counter-terms that respect the original structure of the theory. An interesting aspect of the method is that it allows the proper handling of the disturbing $\mu_I$-dependent ultraviolet divergence. The one-loop constituent quark contributions are treated in the approximation whereby $\mu_I^2$ is taken as small compared to $g^2(v^2+\Delta^2)$ and working up to ${\mathcal{O}}(\mu_I^2)$. After determining the model parameters, we have studied the evolution of the chiral and isospin condensates as well as the pressure, energy and isospin densities and the sound velocity. We have compared the model results with a recent NJL calculation of the same quantities and with LQCD data. The model does a good description for $\mu_I\lesssim 1.5\ m_0$, except perhaps for the sound velocity for which it does not reproduce the peak seemingly appearing in the LQCD calculations.

The results are encouraging and set the stage to explore whether the method can be used to incorporate the effect of meson fluctuations. The method also lends itself to include in the description higher powers of $\mu_I^2$ as well as finite temperature effects. We are currently exploring these avenues and will report on the findings elsewhere in the near future.

\section*{Acknowledgments}
The authors are grateful to G. Endr\"odi and B. B. Brandt for kindly sharing their LQCD data in tabular form. Support for this work was received in part by UNAM-PAPIIT IG100322 and by Consejo Nacional de Ciencia y Tecnolog\'ia grant number A1-S-7655. L. A. H. acknowledges support from a PAPIIT-DGAPA-UNAM fellowship. This work was partially supported by Conselho Nacional de Desenvolvimento Cient\'ifico 
e Tecno\-l\'o\-gico  (CNPq), Grant No. 309598/2020-6 (R.L.S.F.); 
Funda\c{c}\~ao de Amparo \`a Pesquisa do Estado do Rio 
Grande do Sul (FAPERGS), Grants Nos. 19/2551- 0000690-0 and 19/2551-0001948-3 (R.L.S.F.). A.B. acknowledges the support from the Alexander von Humboldt Foundation postdoctoral research fellowship in Germany. J.L.H. acknowledges support from a research assistantship granted by CONACyT-SNI, from the program Unidad de Excelencia María de Maeztu CEX2020-001058-M and from the project PID2019-110165GB-I00PID, financed by the Spanish MCIN/ AEI/10.13039/501100011033/.

\appendix

\section{One-loop constituent quark contribution to the effective potential}
\label{apndx_A}

The thermodynamic potential accounting for the constituent quark contribution at one-loop order is given by 
\begin{equation}
V^1_{f} = iV^{-1} \ln (\mathcal{Z}_{ f}^1),
\end{equation}
where
\begin{equation}
\ln{(\mathcal{Z}_{ f}^1)}=\ln{\left(\det{\left(S_{\rm mf}^{-1} \right)}\right)},
\end{equation}
and $V$ is the space-time volume. Also here, $S_{\rm mf}^{-1}$ is the inverse propagator of the two light-constituent quark species.
Therefore, we are bound to compute the determinant of a matrix $M$ of the form
\begin{equation}
M= \begin{pmatrix}
A & B \\
C & D
\end{pmatrix},
\end{equation}
where  $A,\, B,\, C,\, D$ can be thought of as $p\times p$, $p\times q$, $q\times p$, and $q\times q$ complex matrices, respectively. When $A$, and $D$, are invertible, the determinant of $M$ is given by 
\begin{eqnarray}
\det{(M)}=\det{(A)}\det{(D-CA^{-1}B)}, \label{Eq.detM1}\\
\det{(M)}=\det{(D)}\det{(A-BD^{-1}C)}.\label{Eq.detM2}
\end{eqnarray}
Equation~\eqref{Eq.detM1} can be written as
\begin{equation}
\begin{split}
\det{(M)}&=\det{(A)}\det{(D-CA^{-1}B)}\\
&=\det{(A)}\det{(C^{-1}C)}\det{(D-CA^{-1}B)} \\
&=\det{(-C^{2}A^{-1}BC^{-1}A+CDC^{-1}A)},
\end{split}
\label{detD1}
\end{equation}
whereas Eq.~\eqref{Eq.detM2} as
\begin{equation}
\begin{split}
\det{(M)}&=\det{(D)}\det{(A-BD^{-1}C)} \\
&=\det{(D)}\det{(C^{-1}C)}\det{(A-BD^{-1}C)} \\
&=\det{(-CB+CAC^{-1}D)}.
\end{split}
\label{DetD2}
\end{equation}
For our purposes, $B=C=ig\Delta \gamma^{5}$. Thus, from Eqs.~\eqref{detD1}
and~\eqref{DetD2}, we obtain
\begin{eqnarray}
\det{(M)}= \det{(-C^{2}+CDC^{-1}A)}, \\
\det{(M)}= \det{(-C^{2}+CAC^{-1}D)}. 
\end{eqnarray}
We explicitly compute both expressions. Fist, we use that the standard spin projectors $\Lambda_\pm$ satisfy
\begin{equation}
\gamma^{0}\Lambda_{\pm}\gamma^{0}= \tilde{\Lambda}_{\mp},
\end{equation}
and
\begin{equation}
\gamma^{5}\Lambda_{\pm}\gamma^{5}= \tilde{\Lambda}_{\pm},
\end{equation} 
with the projectors $ \tilde{\Lambda}_{\pm}$ defined as
\begin{equation}
 \tilde{\Lambda}_{\pm} =\frac{1}{2} \left(1\pm \frac{\gamma^0 (\vec{\gamma} \cdot \vec{k}-gv)}{E_k}\right).
\end{equation}
Next, we notice that $A=S_u^{-1}$ and $D=S_d^{-1}$. Therefore, working first in the absence of an isospin chemical potential, for which
\begin{eqnarray}
S_u^{-1}=S_d^{-1}=k_0\gamma^0-\vec{\gamma}\cdot\vec{k}-gv,
\label{nomuI}
\end{eqnarray}
\begin{equation}
\begin{split}
D_1&\equiv -C^{2}+CDC^{-1}A\\
&= g^2\Delta^2+(ig\Delta \gamma^5) S_{d}^{-1} \left(\frac{1}{ig\Delta} \gamma^5 \right)  S_{u}^{-1} \\
&= g^2\Delta^2-\left[k_0^{2}-\left(E_k^u\right)^{2} \right] \Lambda_{-} -\left[k_0-\left(E^d_k\right)^{2}\right]\Lambda_+ , \\
\end{split}
\label{D1}
\end{equation}
and
\begin{equation}
\begin{split}
D_{2}&\equiv -C^{2}+CAC^{-1}D\\
&= g^2\Delta ^{2}+ \gamma^5 S_{u}^{-1} \gamma^5 S_{d}^{-1} \\
&=g^2\Delta^{2}-\left[k_0^{2}-\left(E_k^d \right)^{2} \right]\Lambda_{-} -\left[k_0^{2}-\left(E_k^u \right)^{2} \right] \Lambda_{+}.
\end{split}
\label{D2}
\end{equation}
Thus, using that $\Lambda_+ +\Lambda_{-}=\unity$ and defining $E^{q}_{\Delta}=\sqrt{\left( E^{q}_{k}\right)^{2}+g^2 \Delta^2}$, we have
\begin{eqnarray}
\!\!\!\!D_1&=& - \left( k_0^{2}-\left( E_\Delta^u \right)^{2} \right) \Lambda_{-} -\left( k_0-\left(E^d_\Delta \right)^{2} \right) \Lambda_+ , \,\,\,\,\,\, \\
\!\!\!\!D_2&=& - \left( k_0^{2}-\left(E_\Delta^d \right)^{2} \right) \Lambda_{-} -\left(k_0^{2}-\left(E_\Delta^u \right)^{2} \right) \Lambda_{+} ,  
\end{eqnarray}
and
\begin{equation}
\det{(S_{\rm mf}^{-1})}=\det{(D_1)}=\det{(D_2)}.
\end{equation}
Note that 
\begin{equation}
\begin{split}
\ln{(\mathcal{Z}_{f}^{1})} &=\ln{\left(\det{\left( S_{\rm mf}^{-1} \right)}\right)}\\
&=\frac{1}{2}\ln{\left(\det{\left( S_{\rm mf}^{-1} \right)^2}\right)}\\
&=\frac{1}{2} \ln{\left(\det{\left(D_1 D_2 \right)}  \right)}\\
&=\frac{1}{2} \text{Tr}\left[\ln{\left( D_1 D_2 \right)}  \right],
\end{split}
\end{equation}
and since the product $D_1D_2$ is given by
\begin{equation}
D_1D_2=\left( k_0^{2}-\left( E_\Delta^u \right)^{2} \right) \left( k_0^{2}-\left(E_\Delta^d \right)^{2} \right),
\end{equation}
we get
\begin{equation}
\ln{(\mathcal{Z}_{f}^{1})}=\frac{1}{2}\sum_{q=u,d}  \text{Tr}\left[\ln{\Big( k_0^{2}-\left( E_\Delta^q \right)^{2} \Big)}  \right],
\end{equation}
where the trace is taken in Dirac, color (factors of $4$ and $N_c$, respectively), and in coordinate spaces, namely,
\begin{equation}
\begin{split}
\ln{(\mathcal{Z}_{f}^{1})}&=2N_c \sum_{q=u,d} \int d^{4}x \Big\langle x \Big\lvert \ln{ \Big( k_0^{2}-\left( E_\Delta^q \right)^{2} \Big) } \Big\rvert x \Big\rangle \\
&=2N_c \sum_{q=u,d} \int d^{4}x \int \frac{d^{4}k}{(2\pi)^4}  \,  \ln{ \Big( k_0^{2}-\left( E_\Delta^q \right)^{2} \Big)}.
\end{split}
\end{equation}
Therefore
\begin{equation}
\ln{(\mathcal{Z}_{f}^{1})}=2V N_c \sum_{q=u,d}  \int \frac{d^{4}k}{(2\pi)^4}  \,  \ln{ \Big( k_0^{2}-\left( E_\Delta^q \right)^{2} \Big)  }.
\end{equation}
In order to obtain a more compact expression, we integrate and differentiate with respect to $E_{\Delta}^{q}$ as follows 
\begin{equation}
\ln{(\mathcal{Z}_{f}^{1})}=2V N_c \sum_{q=u,d}  \int \frac{d^{4}k}{(2\pi)^4} \int dE_{\Delta}^q \, \frac{E_{\Delta}^q}{k_0^2-(E_{\Delta}^q)^2}.
\label{intdiff_fermion}
\end{equation}
Performing a Wick rotation $k_0 \to ik_4$, we obtain
\begin{equation}
\ln{(\mathcal{Z}_{f}^{1})}=4iV N_c \sum_{q=u,d}  \int \frac{d^{4}k_E}{(2\pi)^4} \int dE_{\Delta}^q \, \frac{E_{\Delta}^q}{k_0^2-(E_{\Delta}^q)^2},
\label{wickrot_fermion}
\end{equation}
and integrating over $k_4$ and $E_{\Delta}^q$, in this order, we get
\begin{equation}
\ln{(\mathcal{Z}_{f}^{1})}=2iV N_c \sum_{q=u,d}  \int \frac{d^{3}k}{(2\pi)^3}\, E_{\Delta}^q,
\end{equation}
with $\Re[(E_{\Delta}^q)^2]\geq 0$.
Therefore, the constituent quark contribution to the effective potential at one-loop order is given by
\begin{equation}
    V_{f}^{1}=iV^{-1}\ln{(\mathcal{Z}_{f}^{1})}.
\end{equation}
Thus,
\begin{equation}
     V_{f}^{1}=-2 N_c \sum_{q=u,d}  \int \frac{d^{3}k}{(2\pi)^3}\, E_{\Delta}^q.
\label{quarkconteffpot}
\end{equation}
In the presence of an isospin chemical potential for which
\begin{eqnarray}
S_u^{-1}&=&(k_0+\mu_I)\gamma^0-\vec{\gamma}\cdot\vec{k}-gv,\nonumber\\
S_d^{-1}&=&(k_0-\mu_I)\gamma^0-\vec{\gamma}\cdot\vec{k}-gv,
\label{nomuI}
\end{eqnarray}
and repeating the steps starting from Eq.~(\ref{D1}), we obtain Eq.~(\ref{quarkconteffpot}), with the energies $E_\Delta^u$ and $E_\Delta^d$ given by Eqs.~(\ref{fermionenergies}).

We now proceed to the explicit computation of Eq.~(\ref{fromfermions}). In the limit where $\mu_I^2/[g^2(v^2+\Delta^2)]$ is small, Eq.~(\ref{quarkconteffpot}) can be written as in Eq.~(\ref{V1finf}). We use dimensional regularization. The first of the integrals on the right hand side of Eq.~(\ref{V1finf}) is expressed as
\begin{eqnarray}
\int\frac{d^3k}{(2\pi)^3}\sqrt{k^2+g^2v^2+g^2\Delta^2}&\to&
\Lambda^{3-d}
\frac{\Gamma\left(-\frac{1}{2}-\frac{d}{2}\right)}{(4\pi)^{\frac{d}{2}}\Gamma\left(-\frac{1}{2}\right)}\nonumber\\
&\times&\left(\frac{1}{g^2v^2+g^2\Delta^2}\right)^{-\frac{1}{2}-\frac{d}{2}}.\nonumber\\
\label{firstdivint}
\end{eqnarray}
Taking $d\to 3-2\epsilon$ and working in the $\overline{\mbox{MS}}$ scheme
\begin{eqnarray}
\Lambda^2\to \frac{\Lambda^2e^{\gamma_E}}{4\pi},
\end{eqnarray}
where $\gamma_E$ is the Euler-Mascheroni constant, we get
\begin{widetext}
\begin{eqnarray}
\int\frac{d^3k}{(2\pi)^3}\sqrt{k^2+g^2v^2+g^2\Delta^2}&\to&
-\frac{(g^2v^2+g^2\Delta^2)^2}{2(4\pi)^2}\left[\frac{1}{\epsilon}+\frac{3}{2}+\ln\left(\frac{\Lambda^2}{g^2v^2+g^2\Delta^2}\right)\right].
\label{firstdivintexpl}
\end{eqnarray}
The second of the integrals on the right hand side of Eq.~(\ref{V1finf}) is expressed as
\begin{eqnarray}
\int\frac{d^3k}{(2\pi)^3}\frac{1}{(k^2+g^2v^2+g^2\Delta^2)^{3/2}}\to
\Lambda^{3-d}
\frac{\Gamma\left(\frac{3}{2}-\frac{d}{2}\right)}{(4\pi)^{\frac{d}{2}}\Gamma\left(\frac{3}{2}\right)}\left(\frac{1}{g^2v^2+g^2\Delta^2}\right)^{\frac{3}{2}-\frac{d}{2}}.\nonumber\\
\label{secdivint}
\end{eqnarray}
Taking $d\to 3-2\epsilon$ and working in the $\overline{\mbox{MS}}$ scheme we get
\begin{eqnarray}
\int\frac{d^3k}{(2\pi)^3}\frac{1}{(k^2+g^2v^2+g^2\Delta^2)^{3/2}}\to
\frac{2}{(4\pi)^2}\left[\frac{1}{\epsilon}+\ln\left(\frac{\Lambda^2}{g^2v^2+g^2\Delta^2}\right)\right],
\label{secdivintexpl}
\end{eqnarray}
from where the result of Eq.~(\ref{explVdiv}) follows.
\end{widetext}


\bibliography{bibliography}

\end{document}